# Crystal Structure-Based Multioutput Property Prediction of Lithium Manganese Nickel Oxide using EfficientNet-B0

*Chee Sien Wong[#], Benediktus Madika[#], Jiwon Yeom, Youngwoo Choi, Seungbum Hong\**

Department of Materials Science and Engineering, Korea Advanced Institute of Science and Technology (KAIST), 291 Daehak-ro, Yuseong-gu, Daejeon, 34141, Republic of Korea

[#]These authors contributed equally

*Corresponding author: seungbum@kaist.ac.kr

**Abstract**: Here, we present an EfficientNet-B0-based model to directly predict multiple properties of lithium manganese nickel oxides (LMNO) using their crystal structure images. The model is supposed to predict the energy above the convex hull, bandgap energy, crystal systems, and crystal space groups of LMNOs. In the last layer of the model, a linear function is used to predict the bandgap energy and energy above the convex hull, while a SoftMax function is used to classify the crystal systems and crystal space groups. In the test set, the percentages of coefficient of determination ($R^2$) scores are 97.73% and 96.50% for the bandgap energy and energy above the convex hull predictions, respectively, while the percentages of accuracy are 99.45% and 99.27% for the crystal system and crystal space group classifications, respectively. The class saliency maps explain that the model pays more attention to the shape of the crystal lattices and gradients around the lattice region occupied by the larger ions. This work provides new insight into using an intelligent model to directly relate the crystal structures of LMNO materials with their properties.

1. Introduction

Advancements in lithium-ion battery (LIB) technology, mainly through the development of battery materials, have significantly contributed to the progress in electric vehicle and portable electronic device development [1]. The choice of material for the positive electrode in LIBs is critical, influencing voltage and energy storage capacity. $Li_xMn_{2-z}Ni_zO_4$ (LMNO)-based electrodes are gaining prominence over the traditional $LiCoO_2$-based electrodes due to their superior energy density, affordability, extended cycle life, and enhanced safety features [2]. However, the presence of $Mn^{3+}$ in LMNO induces Jahn–Teller distortions, which further affect the electrochemical stability of the material by altering the crystal structure [3]. Consequently, a thorough understanding of the structure of LMNO is crucial for devising effective solutions to its stability challenges.

Despite the potential of using LMNO as a positive electrode in LIBs, correlating its crystal structure at the atomic scale with its properties remains challenging. Transmission electron microscopy (TEM) can visualize the atomic-scale structure of materials, providing detailed insights into their structural information. However, TEM falls short of directly correlating these structures with specific material properties, such as stability and electronic properties. This limitation poses a considerable challenge in the process of material optimization, highlighting the need for integrative approaches to bridge the gap between structural analysis and functional properties.

Machine learning (ML) is a potential solution to the abovementioned challenge. ML algorithms can use material structures obtained from TEM to identify crystal structures and orientations and relate them with material properties [4]. This insight opens an opportunity to predict critical properties of LMNO, such as bandgap energy ($E_{gap}$), energy above the convex hull ($E_{hull}$), crystal system, and crystal space groups. From a battery material standpoint, it is essential to examine $E_{gap}$ and $E_{hull}$ of LMNO since $E_{gap}$ is strongly linked to the electronic conductivity and rate capability of cathode materials [5], while $E_{hull}$ is relatable to the decomposition energy of the compound into a linear combination of stable phases [6]. Additionally, classifying crystal systems and space groups is incredibly useful in analyzing the physical properties of materials and can facilitate the quick screening of potential materials [7].

The reason why ML can be an effective tool for discovering the relationship between crystal structures and LMNO properties is that ML can decipher complex features from intricate data, such as the arrangement and orientations of atoms in the crystal structure of materials [8]. Beyond just structural analysis, ML has a proven track record of becoming a widely accepted approach in materials science, particularly for accelerating materials design and discovery using experimental and computational data [9]. Its utility spans across characterizing materials, predicting molecular properties, enhancing simulation speeds, and facilitating the discovery of new materials [10].

To address the challenge of directly linking the structure and properties of materials using randomly oriented atomic-scale structural information, we propose an ML approach utilizing the diverse orientations and structural appearances of randomly oriented crystal structures of LMNO simulated in VESTA software [11], which can simulate the structural information obtained from real-space TEM images. Using the simulated crystal structures, we utilized an EfficientNet-B0-based ML to simultaneously predict $E_{gap}$, $E_{hull}$, crystal system, and crystal space groups of LMNOs. This multioutput property prediction, a cornerstone of the methodology in this work, encompasses simultaneous predictions of electronic and structural properties, which are crucial for understanding the performance of LMNO when utilized as a positive electrode in LIBs. Hence, this approach offers a broader understanding of the LMNO properties predicted using the crystal structure images, delivering a more cohesive view of the LMNO evaluation for use in LIB applications.

All LMNO structures and their related properties to be predicted were obtained from the materials project [12]. The retrieved structures are in the form of crystallographic information files (CIFs). Using VESTA software, the CIFs were converted into space-filling style crystal structure images and augmented by randomly rotating them to simulate different viewpoints, like in the case of randomly chosen TEM images with unknown crystallographic directions. Data augmentation was conducted to help increase the diversity and variability of the limited data, allowing learning models to learn patterns from the data. However, it is essential to note that the original and augmented images share the same target data in augmentation [13]. Finally, we used the images to train, validate, and test our EfficientNet-B0-based model [14], which had been designed to contain three activation functions in its last layer, hence enabling us to simultaneously predict $E_{hull}$ and $E_{gap}$ via a linear activation function and classify the crystal system and crystal space group via a SoftMax activation function.

## 2. Methods

*2.1 Data Acquisition and Preparation*

All LMNO crystal structures and their related properties to be predicted were retrieved from the materials project database [12]. There were 60 LNMO crystal structures found in the database with seven types of crystal systems and 19 space groups; $E_{gap}$ ranged from 0 eV to 1.6272 eV, and $E_{hull}$ ranged from 0 eV to 0.8195 eV.

We created randomly rotating LMNO crystal structures in the Vesta software [11] to perform image augmentation. In this case, each LMNO crystal structure was opened in the CIF format, and then the animation tool was used to randomly rotate it in a space-filling style. While doing so, PyAutoGUI [15] was run and recorded the randomly rotating crystal structure, with a time interval and a bounding box to generate videos. Finally, the openCV-python [21] was used to create 60 crystal structure images with 512×512×3 pixels from the videos. The images were further compressed to 224×224×3 pixels and normalized by a factor of 255 before being used as the input dataset for the model prediction. The total number of LMNO data points in the dataset was 3,360.

*2.2 Training, Validation, and Testing of the Model*

A 15% test set (549 data points) was created from the original dataset (3,360). The remaining dataset was split into a 20% validation set (778 data points) and an 80% training set (2,333 data points), which were used to train and validate the EfficientNet-B0-based model. The model included three activation functions: one linear and two SoftMax functions in its last layer. The linear function predicted $E_{gap}$ and $E_{hull}$ via multioutput regression, while the SoftMax functions classified the crystal system and crystal space group separately. The mean squared error (MSE) and mean absolute error (MAE) were used to evaluate the regression task, while categorical cross-entropy loss functions and accuracy metrics were used to evaluate the classification tasks. The model used an Adam optimizer with a learning rate of 0.001 and a batch size of 16. Finally, early stopping was applied with a patience of 30 to stop training and validation whenever the model achieved its best performance at a specific epoch.

*2.3 Saliency Mapping*

The saliency map from the Keras-vis package [22] was used to reveal the attention of the model on the LMNO crystal structure images. We only investigated the model attention on the LMNO crystal structure for LMNO crystal system classification. In this case, we picked one crystal structure image belonging to each crystal system for prediction and then used saliency map parameters to generate the mapped images; for example, we used ReplaceToLinear as the model modifier, a smooth noise of 0.05, and a gray color map. To quantify the similarity of the saliency-mapped images, we also calculated their similarity indices with the structural similarity index (SSM) method provided by scikit-image. Before doing so, we transformed and resized the images to 256 ×256.

## 3. Results and discussion

*3.1 Randomly Oriented LMNO Crystal Structures and EfficientNet-B0 Model*

**Fig. 1** demonstrates the process of augmenting the randomly oriented LMNO crystal structures. To do this, each LMNO CIF was opened as a video in Vesta software and then rotated randomly with a space-filling style. Simultaneously, the PyAutoGUI [15] captured the moving video to produce 60 randomly oriented crystal structure images for each LMNO. The generated image came with a bounding box (black lines in the images) and a background color. By examining the images, it is apparent that the different atoms have different sizes and colors, while the same atoms have the same size and color. These distinguishable crystal orientations became the unique features of each LMNO crystal structure from which the model learned to perform predictions.

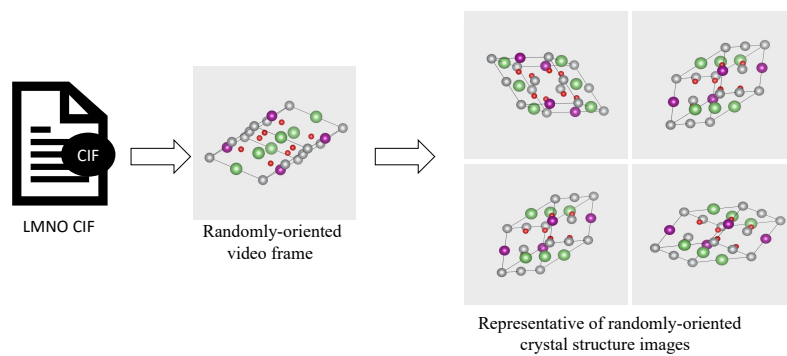

**Fig. 1.** The process of generating the randomly oriented crystal structures of LMNOs. First, each CIF of LNMO structures was opened using Vesta software as a video. At the same time, the PyAutoGUI captured 60 randomly oriented crystal structure images from the video.

To prepare the images for the learning model, they were first uniformly compressed down to a size of 224×224×3 and then scaled by a factor of 225. Then, they were converted into NumPy arrays of numbers before being input into the model, as illustrated in **Fig. 2**. The pre-trained EfficientNet-B0 was used as the ML model, which is based on the mobile inverted bottleneck building block and developed through a neural architecture search framework that considers certain constraints when finding the best architecture in the search space architecture [14,16]. In this model, the final layer uses a specific activation function to make predictions. This work used the linear activation function to predict $E_{gap}$ and $E_{hull}$ and the SoftMax activation function to predict crystal systems and crystal space groups of LMNOs.

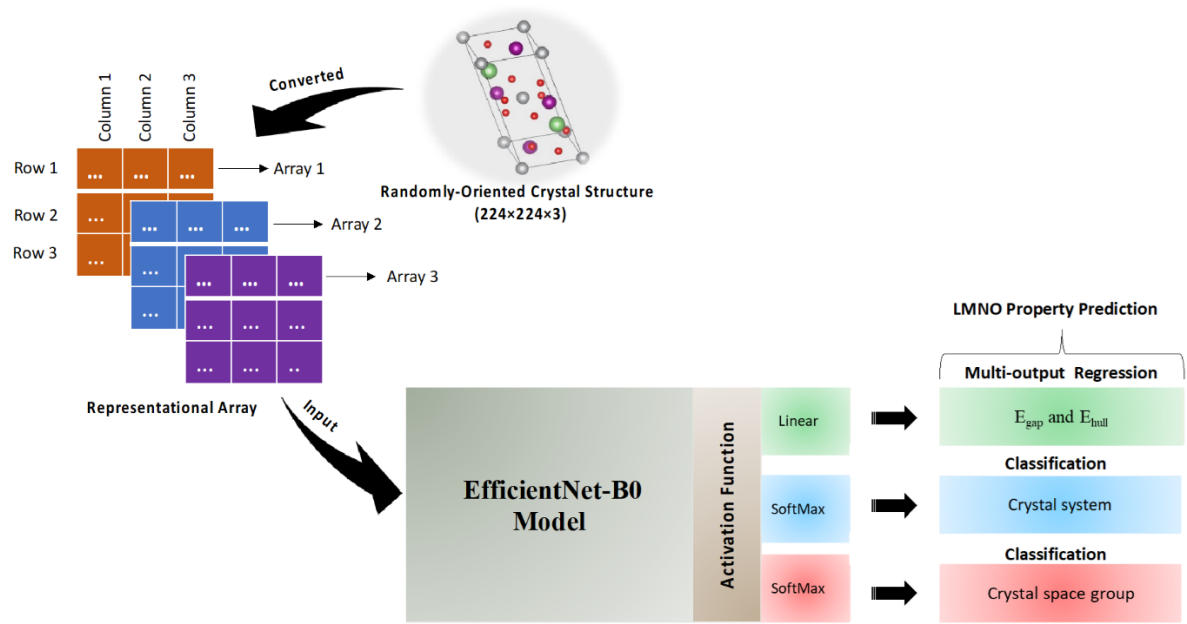

**Fig. 2.** The images were compressed to 224×224×3, converted into NumPy arrays, and finally used as input into the EfficientNet-B0-based model. The last layer uses two activation functions: a linear function to predict $E_{gap}$ and $E_{hull}$ and a SoftMax function to predict crystal systems and space groups of LMNOs.

The model network utilized convolutional operations, scaling, and global pooling on the input data before sending it to the activation functions for making predictions. An early stopping parameter was applied during training and validation to avoid overfitting and long training times. The process was terminated after reaching convergence at the 124th epoch, as shown in **Fig. S1**.

*3.2 Multioutput Regression to Predict $E_{gap}$ and $E_{hull}$ of LMNO*

During the training and validation, the model adjusted the weight and bias by minimizing the difference between the predicted and actual values of $E_{gap}$ and $E_{hull}$ via the mean squared error (MSE) loss function. Loss function minimization was continuously performed until a satisfactory performance level was reached on the training and validation dataset or when convergence to a local minimum was achieved. In **Fig. S1**a, the training and validation MSEs drastically drop to low values after a few epochs and gradually decrease until they almost become constant at approximately 0. However, the validation curve is less stable and higher than the training curve. The mean absolute error (MAE) curves in **Fig. S1**b also show a similar trend after a few epochs, but they still decrease as the number of epochs increases. The final MAE reaches a value of approximately 0.1. However, the validation curves are less stable and higher than the training curve, suggesting that the validation performances are still not as good as the training performance.

Another way to evaluate model performance is through a parity plot, a representation of the predicted values of a trained model compared to the actual values. The plot can help visually show how well a model fits the data and detect patterns or trends. The parity plots for validation and testing are depicted in **Fig. 3**a for $E_{gap}$ and **Fig. 3**b for $E_{hull}$. The trend of the validation and test plots tends to be the same as that of the ideal predictions, increasing from the lower value to the higher value of $E_{gap}$ and $E_{hull}$. The model prediction errors in the test set are almost consistent since deviations from both sides (below and above) of the ideal line tend to have equal magnitude, meaning that the model does not have a high bias toward overestimating or underestimating the predictions in the test set. However, there are some more significant deviations from the ideal line in the case of the validation set prediction, especially for the higher valued-$E_{gap}$ and $E_{hull}$. This might be due to the nature of the target data distribution. In this case, fewer data points exist in the higher-valued range than in the lower-valued range of $E_{gap}$ and $E_{hull}$, as shown in **Fig. S2**. Such a case can make the model learn only a little information in the higher-valued ranges but also rich information from the lower-valued ranges. Overall, the high performance of the model can be judged from the high percentages of the $R^2$ score test: 97.73% for $E_{gap}$ and 96.50% for $E_{hull}$.

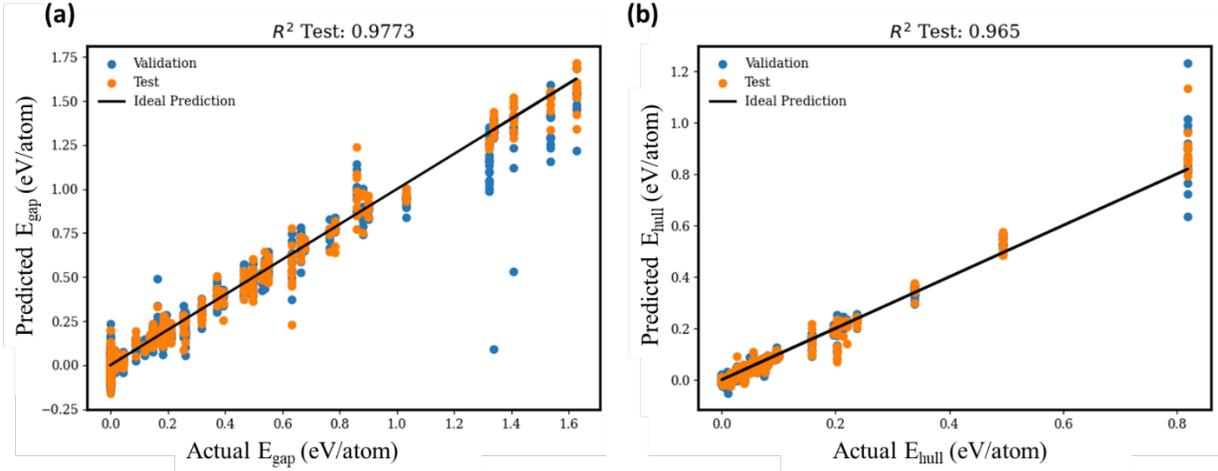

**Fig. 3.** Training and test parity plots with $R^2$ scores of (a) $E_{gap}$ and (b) $E_{hull}$.

*3.3 Classification of the Crystal System and Crystal Space Groups of LMNO*

The working mechanism and the model architecture of the EfficientNet-B0 to classify crystal systems and crystal space groups are the same as the regression above. However, the only difference is that the classification task used the SoftMax function as the activation function. Usually, the SoftMax function, which outputs values from 0 to 1, is used in the multiclass task, where it returns probabilities of each class, with the highest probability corresponding to the target class [17]. In this work, the SoftMax function mapped the randomly oriented crystal structure images that EfficientNet-B0 had processed to a probability distribution over crystal systems and crystal space groups.

The loss curves, which are categorical cross-entropies, and accuracies for the training and validation are shown in **Fig. S1**c-d for the crystal system classification and in **Fig. S1**e-f for the crystal space group classification of LMNOs. The training and validation losses drop drastically to approximately 0 after a few epochs and remain almost constant. This makes the accuracy curves for training and validation drastically increase to values close to 1.0 after a few epochs and remain practically constant. However, the validation loss and accuracy curves are unstable because they decrease and increase significantly at certain epochs.

To assess the accuracy of the model classification of data, confusion matrices were created for each classification performance based on the number of true positive (TP), true negative (TN), false positive (FP), and false negative (FN) samples [18]. These matrices are displayed in **Fig. 4**a-b for the test set consisting of 549 data points. This shows that almost all crystal systems and space groups are correctly classified in both cases. The model misclassifies only three triclinic crystals into three cubic crystals, as shown in **Fig. 4**a.

Meanwhile, in **Fig. 4**b, one P2_13, one P2_1, and two P-1 crystal space groups are misclassified into one P4_332, one P-1, and two P4_332 groups, respectively. In addition, the confusion matrices were plotted for the validation sets of crystal systems and crystal space groups, as shown in **Fig. S2**a-b, respectively. Four triclinic crystals are misclassified into 4 four cubic crystals, as shown in **Fig. S2**a. Meanwhile, one P2_13, three P-1, one Cmce, and two C2/c are classified into one P-1, three P4_332, one C2/m, and one Cmce and one C2/m, respectively. The same crystal system is misclassified into the wrong one in the validation and test sets. The model consistently classifies the LMNO crystal systems in different datasets. However, some misclassified crystal space groups are not the same in the test and validation sets in the LMNO crystal space group classification. This indicates that the model is less consistent in classifying the LMNO crystal systems in different datasets. Overall, the model demonstrates a high accuracy in classifying the test set, indicating that the model accurately predicted the correct labels for most instances in each crystal class [19].

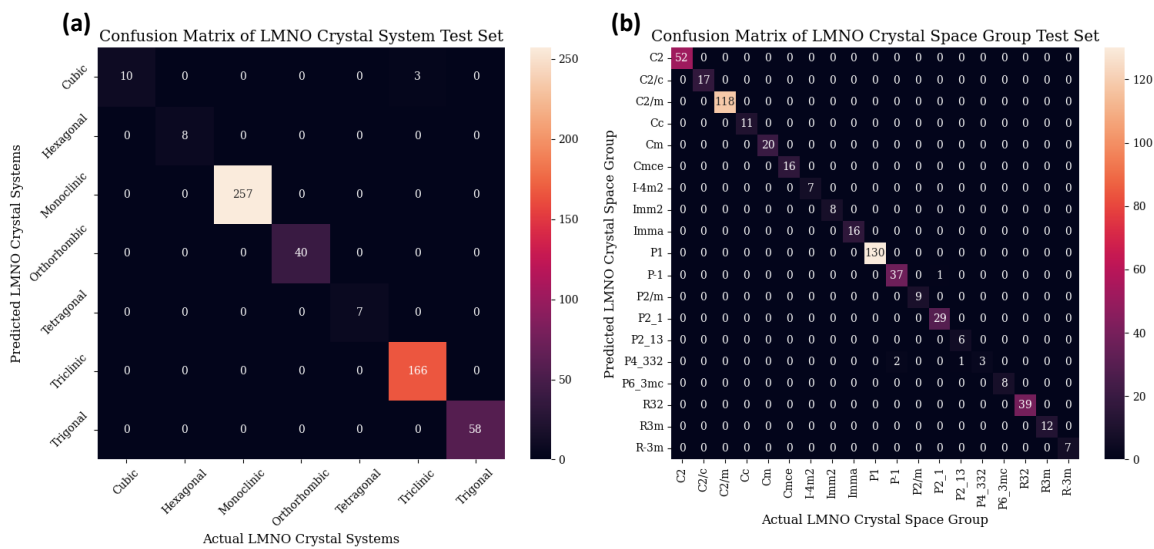

**Fig. 4.** Confusion matrices in test sets of (a) crystal system and (b) crystal space group classifications.

The classification report for crystal system and crystal space group classifications in the test sets are shown in **Table 1** and **Table 2**, respectively. The table includes precision, recall, F1-score, and several support data points. Precision and recall are determined at a single point, reflecting the model performance at a specific threshold. The classification reports quantify the prediction presented in the confusion matrices above. In **Table 1**, only the triclinic and cubic crystals have F1-scores below 1.000. This is in line with the misclassification observed in the confusion matrix of the crystal system above, where three triclinic crystals are misclassified into three cubic crystals. Therefore, the triclinic crystal has a precision of 0.9822, while the cubic crystal has a recall of 0.7692. **Table 2** shows the quantified prediction of the crystal space group. Only P2_13, P2_1, P-1, and P4_332 have F1-scores below 1.000. This is because one P2_13, one P2_1, and two P-1 crystal space groups are misclassified into one P4_332, one P-1, and two P4_332, respectively, as shown in the confusion matrix of the LMNO crystal space group prediction above. P-1, P2_1, and P2_13 have precisions of 0.9487, 0.9667, and 0.8571, respectively, while P-1 and P4_332 have recalls of 0.9737 and 0.500, respectively.

**Table 1.** Classification report of crystal systems in the test set

| Crystal System | Precision | Recall | F1-score | Support |
|---|---|---|---|---|
| Cubic | 1.0000 | 0.7692 | 0.8696 | 13 |

| Class | Precision | Recall | F1-score | Support |
| --- | --- | --- | --- | --- |
| Hexagonal | 1.0000 | 1.0000 | 1.0000 | 8 |
| Monoclinic | 1.0000 | 1.0000 | 1.0000 | 257 |
| Orthorhombic | 1.0000 | 1.0000 | 1.0000 | 40 |
| Tetragonal | 1.0000 | 1.0000 | 1.0000 | 7 |
| Triclinic | 0.9822 | 1.0000 | 0.9910 | 166 |
| Trigonal | 1.0000 | 1.0000 | 1.0000 | 58 |
| Accuracy | | | 0.9945 | 549 |
| Macro average | 0.9975 | 0.9670 | 0.9801 | 549 |
| Weighted average | 0.9946 | 0.9945 | 0.9942 | 549 |

**Table 2.** Classification report of the crystal space groups in the test set

| Class | Precision | Recall | F1-score | Support |
| --- | --- | --- | --- | --- |
| C2 | 1.0000 | 1.0000 | 1.0000 | 52 |
| C2/c | 1.0000 | 1.0000 | 1.0000 | 17 |
| C2/m | 1.0000 | 1.0000 | 1.0000 | 118 |
| Cc | 1.0000 | 1.0000 | 1.0000 | 11 |
| Cm | 1.0000 | 1.0000 | 1.0000 | 20 |
| Cmce | 1.0000 | 1.0000 | 1.0000 | 16 |
| I-4m2 | 1.0000 | 1.0000 | 1.0000 | 7 |
| Imm2 | 1.0000 | 1.0000 | 1.0000 | 8 |
| Imma | 1.0000 | 1.0000 | 1.0000 | 16 |
| PI | 1.0000 | 1.0000 | 1.0000 | 130 |
| P-I | 0.9487 | 0.9737 | 0.9610 | 38 |
| P2/m | 1.0000 | 1.0000 | 1.0000 | 9 |
| P2_1 | 0.9667 | 1.0000 | 0.9831 | 29 |
| P2_13 | 0.8571 | 1.0000 | 0.9231 | 6 |
| P4_332 | 1.0000 | 0.5000 | 0.6667 | 6 |
| P6_3mc | 1.0000 | 1.0000 | 1.0000 | 8 |
| R32 | 1.0000 | 1.0000 | 1.0000 | 39 |
| R3m | 1.0000 | 1.0000 | 1.0000 | 12 |
| R-3 m | 1.0000 | 1.0000 | 1.0000 | 7 |
| Accuracy | | | 0.9977 | 549 |
| Macro average | 0.9947 | 0.9969 | 0.9956 | 549 |
| Weighted average | 0.9980 | 0.9977 | 0.9978 | 549 |

From the confusion matrices and classification reports above, the classes from which the actual data are misclassified have low precision values, while the classes into which the actual data are misclassified have low recall values. The F1-scores are above 0.9 and are as high as 1.00, leading to a total accuracy of 0.9945 (99.45%), a macro average accuracy of 0.9801 (98.01%), and a weighted average accuracy of 0.9942 (99.42%) for the LMNO crystal system classification, and a total accuracy of 0.9977 (99.77%), a macro average accuracy of 0.9969 (99.69%), and a weighted average accuracy of 0.9978 (99.78%) for the LMNO crystal space group classification. The classification reports for the validation sets of the LMNO crystal system and crystal space group classifications are also presented in **Tables** S1-S2, respectively, showing high accuracy. The high accuracy predictions indicate that the model has a high potential to correctly identify positive crystal classes with a low rate of false positives, correctly identify a large percentage of all positive crystal classes, and achieve high precision without sacrificing recall, and vice versa.

Receiver operating characteristic (ROC) curves are constructed to investigate imbalanced classifications. The ROC curves of the crystal system and crystal space group classifications in the test sets are shown in **Fig. 5**a-b, respectively. The ROC curve evaluates the trade-off between the true positive rate (TPR) and the false positive rate (FPR) at different decision thresholds. The TPR (sensitivity or recall) is the ratio of the TP to TP plus FN, indicating the proportion of the actual positive class that the model correctly classifies as positive, while the FPR is the ratio of FP to FP plus TN, determining the proportion of the actual negative class that the model incorrectly classifies as positive. Ideally, the prediction should have a correct positive class prediction rate of 1 (top of the plot) and an incorrect negative class prediction rate of 0 (left of the plot). In other words, the best possible model to achieve perfect prediction is the top-left of the plot, with a coordinate of (0,1)

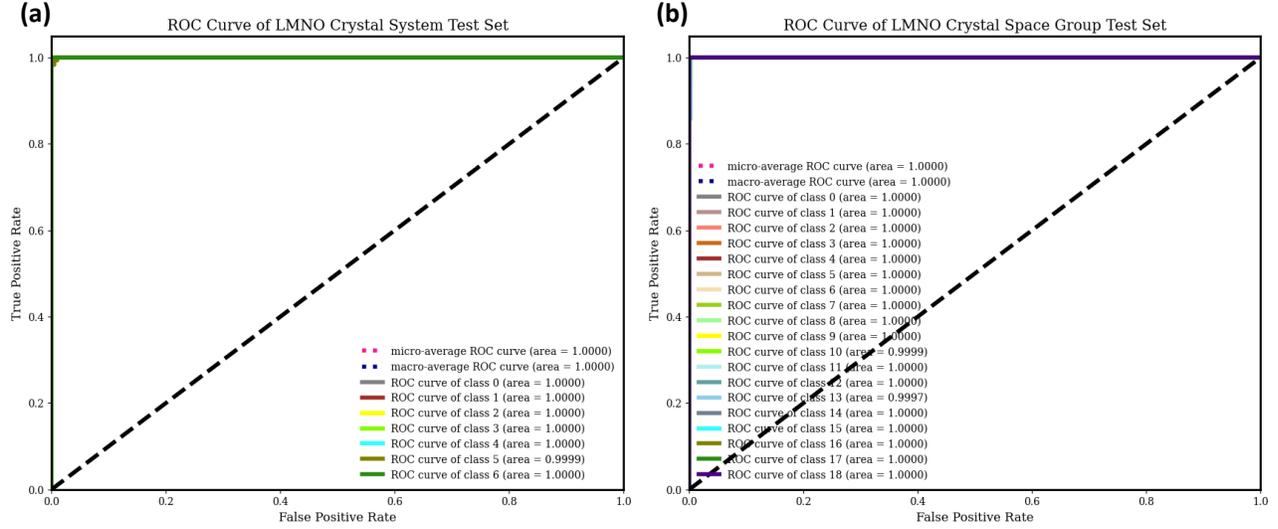

**Fig. 5**. ROC curves in test sets of (a) crystal system and (b) crystal space group classifications of LMNO.

In **Fig. 5**a, the ROC curves of crystal systems are denoted as 0 (cubic), 1 (hexagonal), 2 (monoclinic), 3 (orthorhombic), 4 (tetragonal), 5 (triclinic), and 6 (trigonal). In **Fig. 5**b, the ROC curves of spaces groups are denoted as class 0 (C2), 1 (C2/c), 2 (C2/m), 3 (Cc), 4 (Cm), 5 (Cmce), 6 (I-4m2), 7 (Imm2), 8 (Imma), 9 (P1), 10 (P-1), 11 (P2/m), 12 (P2_1), 13 (P2_13), 14 (P4_332), 15 (P6_3mc), 16 (R32), 17 (R3m), and 18 (R-3 m). The ROC curves in **Fig. 5**a-b are very close to the upper left corner, signifying that the model has a high discriminatory power to distinguish the positive and negative classes of the crystal systems and the crystal space groups, respectively.

This also means that the TPR (sensitivity) is high while the FPR is low, which attests to the fact that the model correctly identifies most positive classes while minimizing the number of false positive classes. Furthermore, the area under the curve (AUC) values (or the area in **Fig**. **5**a of the crystal system classification) is mostly 1.0000. Only the crystal system (triclinic) from which the misclassified data originates has a lower AUC value. The same trend is observed in **Fig. 5**, where only the crystal space groups (P-1 and P2_13) from which the misclassified data originate have AUC values lower than 1.0000. The microaverage ROC and macroaverage ROC are calculated and found to be 1.00. The microaverage ROC focuses on overall performance across classes, while the macroaverage focuses on the performance of the model for each class of crystal systems and crystal space groups. The ROC curves for the LMNO crystal system validation set and crystal space group classifications are also plotted in **Fig. S4**a-b for comparison.

Another performance metric that can investigate imbalanced classification is the precision-recall (PR) curve. **Fig. 6**a-b shows the PR curves for the crystal system and crystal space group classifications in the test sets. The crystal systems in **Fig. 6**a are denoted as 0 (cubic), 1 (hexagonal), 2 (monoclinic), 3 (orthorhombic), 4 (tetragonal), 5 (triclinic), and 6 (trigonal), while the crystal space groups in **Fig. 6**b are denoted as 0 (C2), 1 (C2/c), 2 (C2/m), 3 (Cc), 4 (Cm), 5 (Cmce), 6 (I-4m2), 7 (Imm2), 8 (Imma), 9 (P1), 10 (P-1), 11 (P2/m), 12 (P2_1), 13 (P2_13), 14 (P4_332), 15 (P6_3mc), 16 (R32), 17 (R3m), and 18 (R-3 m). The PR curves of the crystal system and crystal space group classifications in the validation set are also plotted in **Fig. S5**a-b. When the F1 value increases from 0.2 to 0.8, the iso-F1 curve shifts to the right, indicating higher precision at the same level of recall, resulting in a trade-off between higher precision and higher recall and leading to higher average precision (AP) values (1.00).

This suggests that we simply need to obtain an F1 score of 0.8 or above to have a higher threshold of precision and recall simultaneously. High precision and recall are important in classifying the crystal systems and crystal space groups because they can maximize correct identification and minimize incorrect classifications of the model, respectively. The AP values shown in the figures result from integrating the AUC across multiple thresholds, thus representing the summarization of the overall performance of the model across all thresholds by calculating the average precision across the different recall levels.

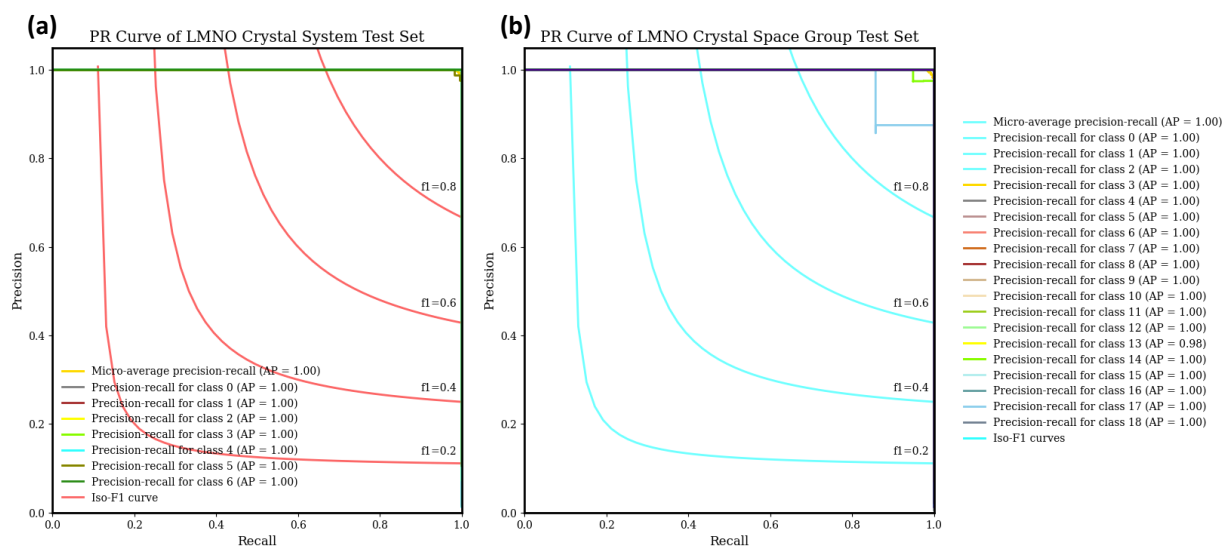

**Fig. 6.** (PR) curve (a) crystal system (b) crystal space group classifications of LMNOs.

*3.4 Visualization of the Crystal System Attention of the EfficientNet-B0 Model*

To understand how the EfficientNet-B0 model received attention from the crystal structures, saliency maps of the crystal system are visualized. A saliency map determines which aspects of the proposed model are most important to obtain accurate predictions. It is calculated by taking the gradient of the model output value concerning the input matrix [20]. The general goal of the saliency map is to transform the original view of an image intuitively, where the high-impact features, such as image pixels and resolutions, are distinguished from the low-impact features with respect to the model prediction outputs.

To reveal the model attention on the input features using the salience map, we selected one of the target predictions to use, namely, the LMNO crystal systems, which are shown in **Fig. 5**. In the figure, the left pictures are the original images, while the right pictures are the salience map images. The model attention, as depicted with high contrast pixels, is concentrated around the areas of interest of the structures. The saliency maps also show more significant gradients around lattice regions occupied by larger ions, which means that the proposed model pays more attention to the spatial arrangement of large ions to capture salient information. This agrees with the general observation that the larger ions form a close-packed structure, whereas the smaller ions occupy the interstices; the close-packed structure determines the crystal structure. Furthermore, the model can also capture the shapes of the crystal structures and use them to make predictions.

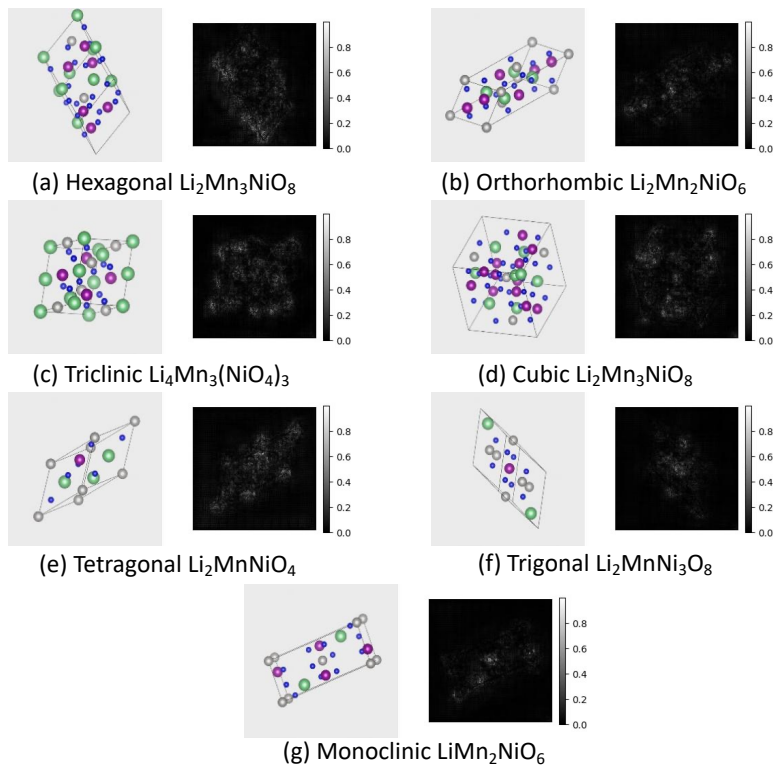

**Fig. 7.** Saliency map of the LNMO crystal systems.

All saliency maps have different patterns; therefore, we can assume they are characteristic of the corresponding crystal systems. This is quantitatively supported by the similarity index values, as shown in **Fig. 8**, which differ for each crystal system. This saliency map, therefore, explains why the proposed EfficientNet-B0 model can result in high-performance predictions.

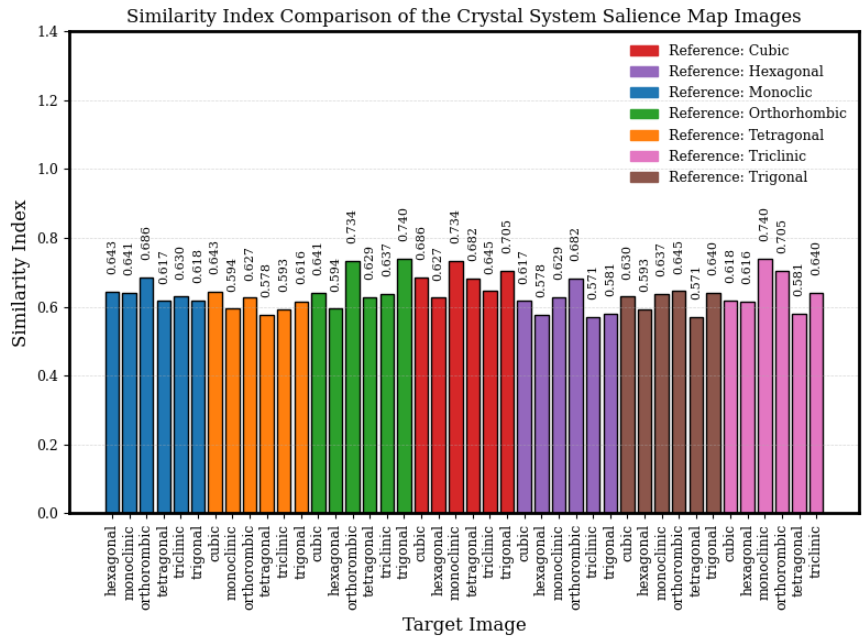

**Fig. 8.** Similarity index comparison of the LMNO crystal system salience map.

Lastly, the model in this work is compared with a crystal graph neural network (CGNN)-based model, which is a more standard approach to predicting material properties based on crystal structures [8]. To make the ML design meet the CGNN model architecture, the CIFs of the LMNOs were translated into a graph format where the atomic positions become nodes in the graph and the interactions (or bonds) between atoms become the edges. The electronic radial distribution function capturing the electronic environment around atoms was used as a global feature. The global features were combined with the atomic number of each atom in the structure to represent the node features of the graph. Meanwhile, each bond in the crystal structure was used to describe the edge of the graph, with the edge feature being the distance between atoms in the crystal structure. The graph data were used as input for the CGNN-based model to make the multi-output property predictions of LMNOs.

The prediction performances of the CGNN-based model are shown in **Fig. S6**. The $R^2$ values of the bandgap and energy above the convex hull fluctuate significantly below 0, as shown in **Fig. S6**a-b, indicating high variance and error. Meanwhile, the train and test set of the crystal system predictions can hit 1.0 (100%) accuracy yet significantly fluctuate, as shown in **Fig. S6**c. For the crystal space group prediction, the train and test accuracy are far separated, where the train accuracy can hit 1.0 (100%) while the test accuracy cannot even reach 0.5, as shown in **Fig. S6**d. From this observation, only the crystal system seems to be predicted well using the CGNN-based models.

## 4. Conclusion

This study represents an advance in addressing the challenge of directly correlating the crystal structure of LMNOs with their electrochemical properties. By leveraging the EfficientNet-B0-based machine learning model, we established a direct, reliable method for simultaneously predicting bandgap energy, energy above the convex hull, crystal systems, and crystal space groups of $Li_xMn_{2-z}Ni_zO_4$ (LMNO) materials from their crystal structure images, which is a limitation of traditional atomic-scale crystal structure visualization techniques, such as transmission electron microscopy. The predictive performances of the model are promising, evidenced by outstanding $R^2$ scores of 97.73% and 96.50% for the bandgap energy and the energy above the convex hull predictions, respectively, and classification accuracies of 99.45% and 99.27% for the crystal systems and the space groups of LMNO. Class saliency maps revealed that the model paid more attention to the larger atoms in the crystal structure images and shapes of the LMNO crystal structures. This study offers a pioneering method that leverages LMNO crystal structure images for multioutput electrochemical property predictions, providing a more integrative analysis and facilitating rapid property predictions using crystal structure images.

**CRediT authorship contribution statement**

**Chee S. Wong**: Writing – original draft, Methodology, Investigation, Formal analysis, Conceptualization, Software, Visualization. **Benediktus Madika**: Writing – original draft, Methodology, Investigation, Formal analysis, Conceptualization, Software, Visualization. **Jiwon Yeom**: Validation, Formal analysis, Writing – review & editing. **Youngwoo Choi**: Validation, Formal analysis, Writing – review & editing. **Seungbum Hong**: Conceptualization, Formal analysis, Project administration, Funding acquisition, Resources, Validation, Writing – review & editing, Supervision. These authors (C.S.W. and B.M.) contributed equally.

**Acknowledgment**

This work is supported by the KAIST-funded Global Singularity Research Program for 2021, 2022 and 2023, and the National Research Foundation of Korea (NRF) grant funded by the Korea government (MSIT) (RS-2023-00247245).

**Declaration of competing interest**

The authors declare that they have no conflict of interest.

**Data availability**

The authors declare that the data supporting this study are available within the article and its supplementary information files or from the corresponding authors upon reasonable request.

**Code availability**

The raw data required to reproduce these findings are available to download from [12]. The processed data and the code required to reproduce these findings are available at https://github.com/MIIMSEKAIST/Li_Mn_Ni_O_crystal_images.

**Declaration of generative AI and AI-assisted technologies in the writing process**

During the preparation of this work the authors used chatGPT 4.0 in order to edit the sentences. After using this tool/service, the authors reviewed and edited the content as needed and take full responsibility for the content of the publication.